% This manuscript is in plain TeX. If you have problems 
% processing it, please contact V. Privman at
% privman@clarkson.edu
\input epsfig

\magnification=\magstep 2

\overfullrule=0pt

\hfuzz=16pt

\baselineskip 16pt

\nopagenumbers\pageno=1\footline={\hfil -- {\folio} -- \hfil}

\

\centerline{{\bf Growth of Monodispersed Colloids by}}
\centerline{{\bf Aggregation of Nucleating Subunits}}

\

\centerline{Jongsoon Park and Vladimir Privman}

\ 
 
\centerline{\sl Department of Physics, Clarkson University}
\centerline{\sl Potsdam, New York 13699--5820, USA}

\ 

\centerline{\sl Electronic mail:\ \ {\tt privman@clarkson.edu}}

\vfill

\noindent{}To be published in the 2000 volume of the series 
{\it Recent Research Developments in Statistical Physics\/}
(Transworld Research Network, Trivandrum, India)

\eject

{\noindent {\bf ABSTRACT} 

  Formation of monodispersed colloidal particles is a
  complex process: nuclei, produced rapidly
  in a supersaturated solution, grow to nanosize primary
  particles, which then aggregate (coagulate) to form much larger final
  colloids. This paper reviews a kinetic model that explains the
  formation of dispersions of narrow size distribution in such processes.
  Numerical simulations of the kinetic equations, with experimental model
  parameter values, are reported. The model was tested for a system
  involving formation of uniform spherical gold particles. The
  calculated average size, the width of the particle size distribution,
  and the time scale of the process, agreed semiquantitatively with the experimental 
values.

\vskip 0.800 true cm

\noindent {\bf 1.\ INTRODUCTION}

We review the recently developed theoretical model [1] of growth
of monodispersed colloids by precipitation from 
homogeneous 
solutions [2-8].
Systematic experimental studies of the mechanisms of synthesis and properties
of such colloids have been initiated about quarter 
a century ago. A large number of uniform dispersions of particles
of various chemical composition and shape, ranging in size from nanometers to few microns, have been described. 
Theoretical description of the formation of uniform colloids relies on experimental identification of solute species that are involved in the various stages of the particle formation and on  observation of growth stages leading from the initial nucleation to final particles. Early theoretical modeling was based on the mechanism suggested by LaMer: a short nucleation burst, followed by diffusional growth of the resulting nuclei to form
identical fine particles [8,9].

However, there has been mounting experimental evidence that the 
burst-nucle\-ation/dif\-fusional growth mechanism alone is inadequate. 
Specifically, it has been found that many spherical particles precipitated
from solution showed polycrystalline X-ray characteristics, such as ZnS
[10], CdS [11], Fe$_2$O$_3$ [12], etc. These particles are not single crystals.
Rather, it has been confirmed by several techniques (small angle light scattering, electron microscopy, X-ray diffraction) that most monodispersed colloids consist of small crystalline subunits [10-20]. Furthermore, it has been observed [1,7,16] that the crystalline subunits in the final particles were of the same size as the diameter of the precursor subunits (singlets) formed in solution, thus suggesting an ag\-gregation-of-subunits mechanism.

The substructure have been identified also in particles of shapes different from spherical, and it has been recognized that different morphologies of the final colloids must be related to the nature of the precursor singlets [12,18-20]. 
This experimental evidence has led to two major theoretical challenges. First, the morphology and shape selection of particles formed by interplay of nucleation and aggregation processes must be explained. Second,
the size-selection mechanism, i.e., the kinetics of generation of narrow particle size distribution, must be identified. Several theoretical approaches utilizing thermodynamic and dynamical growth
mechanisms [1,4,5,18,21-32] have been described in the literature.
Models of aggregation of subunits can be developed that yield a peaked and even sharpening with time particle size distribution [21-34]. However, none of the earlier attempts could fit quantitatively a broad range of experimental findings. Here we review the first results of a new approach [1] that explains the {\it size selection}, by coupling the dynamical equations for the processes of the particle growth by aggregation of subunits and of formation (nucleation) of these subunits.

Thus, the main new finding of our work [1], crucial to obtaining narrow size distribution, has been that the growth of the final, secondary particles by aggregation of singlets, must be coupled, via the rate of formation of these primary particles (singlets), to the time-dependence of the process of their nucleation and growth. We take the simplest possible models of both processes (primary and secondary particle formation). This choice simplifies numerical simulations and thus allows to scan a wider range of parameters. It avoids introduction of unknown microscopic parameters; as a result we only fit {\it one\/} parameter, the effective surface tension, and even that parameter turns out to be close to the experimental bulk value. This review is organized as follows. In Section~2, a kinetic model of secondary particle formation by singlet-capture dominated growth is considered. In Section~3, primary particle (singlet) formation by burst nucleation is incorporated to complete the model. 
Finally, Section~4 reports results of numerical simulations and comparison with an experiment.
 
\vskip 0.800 true cm

\noindent{\bf 2.\ KINETIC MODEL OF COLLOID PARTICLE GROWTH}
 
In colloid particle synthesis of the type considered here, in the initial induction step solutes are formed to yield a supersaturated solution, leading to nucleation. The nuclei might then further grow by a diffusive mechanism. The resulting primary particles (singlets) in turn aggregate to form secondary particles. The latter process, to be modeled in this section, is facilitated by the appropriate chemical conditions in the system: the ionic strength and/or the pH must assume values such that the surface potential approaches the isoelectric point, resulting in reduction of electrostatic barriers, thus promoting particle aggregation. Formation of the final (secondary) particles, which can be of narrow size distribution, is clearly a diffusion-controlled process [1-7].

An important experimentally observed property is that the secondary particles are sufficiently sparsely positioned in solution to consider their evolution as largely independent. Furthermore, we will assume that the particles are spherical, with density close to that of the bulk material, which is experimentally a very common case, and the focus will be on their size distribution due to growth by aggregation, essentially, by consumption of diffusing singlets (mono\-mers, primary particles). 
Several other simplifying assumptions will be made
in order to zero-in on the essential ingredients of the theoretical modeling and make numerical simulations tractable. For instance, we do not address the processes by which the singlets forming secondary particles undergo restructuring and rearrangements resulting in compact structures and leading to shape selection. Some of the approximations made will be discussed at the end of this section.
Thus, we focus on the two main stages in the process: the production of the primary and secondary particles. The latter will be treated presently, while the formation of nuclei and their possible growth (aging) to 
primary particles will be considered in the next section.

Kinetic rate-equation models of aggregation typically utilize a master equation for the distribution of growing particles by their size. Here the size will be defined by how many primary particles (singlets) were aggregated into each secondary particle, denoted by $s=1,2,\ldots$. The growing particles can adsorb or emit singlets and multiplets. The master equation is then quite standard to set up, as in, e.g., [33]. In the present case, the process is experimentally documented to be highly irreversible, so that detachment can be disregarded. By considering the singlet number $s$ as the only parameter characterizing the cluster, we use the approximation that assumes that the internal relaxation/restructuring processes within the cluster (growing secondary particle) are fast, and that they result in an approximately spherical object, of density similar to that of the bulk material, so that the voids have been eliminated and the solvent pushed out in the internal restructuring. We will comment on this point again later.

Furthermore, it is assumed that the singlets have diffusion constants larger than aggregates so that their capture dominates the growth process; 
this approximation will be also discussed again, shortly. The master equation then takes the form
$$ { d  N_s \over  d  t } = w_{s-1} N_{s-1} - w_s N_s \quad (s > 1)\,. \eqno(1) $$
\noindent Here $N_s (t)$ is the time-dependent number density (per unit volume) of the secondary particles consisting of
$s$ primary particles. The attachment rate $w_s$ will be taken from the Smoluchowski expression [35]:
$$ w_s (t) = 4 \pi R_s D N_1 (t) \, , \eqno(2) $$
\noindent where $R_s$ is the radius of the $s$-size particle, given by
$$ R_s = 1.2\, r s^{1/3} \, . \eqno(3) $$
\noindent The parameters $r$, the primary particle radius, and $D$, their diffusion constant, are experimentally available. The constant 1.2
in equation~(3) was calculated as
$$(0.58)^{-1/3}\simeq 1.2 \, , \eqno(4) $$
\noindent where 0.58 is the typical filling factor
of the random loose packing of spheres [36]. The fact that all the primary particles were assumed to have the same radius $r$ will be 
further commented on later.
Note that the rate in equation~(2) involves the product $R_s D \propto rD$, which, according to the Einstein formula for the diffusion constant, is not sensitive to the distribution of values of the radii $r$.

We note that this approach already involves various assumptions. For example,
to derive equation~(2), the attachment process is assumed to be fully irreversible. Furthermore, the above equations are only valid for large $s$. Indeed,
in reality clusters of all sizes $s$ and $S$ can coagulate to form larger clusters of size $s+S$. These are produced with the rate $4 \pi (R_s + R_S)(D_s+D_S)N_s N_S$ within the Smoluchowski approach, where for $s=S$ there is an additional factor of $1/2$ owing to double-counting. Our ignoring all but the singlet capture processes, $S=1$, is supported by the experimental observations but constitutes an approximation. Furthermore, the form (2) is only correct for $s \gg S=1$. Thus, we used the above relations as shown simply to avoid introducing additional parameters. For instance, for $s=1$, this ammounted to ignoring the ``double counting'' factor of $1/2$, and the added factor of 4 due to the radii and diffusion constants of the two singlets adding up.

Ordinarily, the evolution of the population of singlets, which is
not covered by equation (1), is obtained from the conservation of matter:
$$ N_1 (t) + \sum\limits_{j=2}^\infty j N_j(t) = N_1 (0) \, , \eqno(5) $$
\noindent which assumes that initially, at time $t=0$, there are only singlets. When combined with
the assumption of the singlet-capture dominance, the particle distribution is confined to small sizes, as has been 
confirmed by numerical simulations and other studies [1,37]. Indeed, most singlets will simply combine into dimers, fewer trimers, etc., and then the growth stops.
The conventional approach has been to consider more general models, with discrete or continuous population balance, restoring multiplet
aggregation, etc., which adds terms in and modifies the master equation (1); see, e.g., [21-34]. Typically such models have yielded wide
size distributions [33,38-40].

We have developed [1] a new approach based on the observation that the supply of singlets is in itself a dynamical process. Numerical simulations indicate that if the concentration of the singlets were constant, i.e., if they were continuously generated to compensate for their depletion due to aggregation, then the resulting particle size distribution would be wide and peaked at small sizes, such that $N_s \simeq N_1 s^{-1/3}$ for sizes up to some growing cut-off $s$-value.
However, if the supply of singlets is controlled in such a way that their concentration is a decaying function of time, then size selection can be obtained for the secondary particles with some time-dependence protocols. For example, when the rate $\rho (t)$ at which the primary particles are formed (per unit volume) was chosen to be a decaying power-law, then
numerical calculations [1] yielded a single-hump size distribution for the secondary particles. Note that the equation for $N_1(t)$ must be modified by replacing
relation (5) by
$$ N_1 (t) = \int\limits_0^t \rho (t^\prime)\, d t^\prime - \sum\limits_{j=2}^\infty j N_j(t) \, , \eqno(6) $$
\noindent with the initial values $N_s(0)=0$ for all $s=1,2,3,\ldots$. 

Equations of this type, with singlet-capture dominance of the dynamics 
and several ad hoc singlet-input rate functions, have been described in the literature [37]. The emphasis [37] has been on cases which are exactly solvable, e.g., one-dimen\-sional versions, and those where the size distribution shows self-similar behavior.
Our approach is quite different in that we actually {\it model\/} the primary-particle input rate $\rho(t)$ by using the burst-nucleation approach; see the next section.

Let us now further comment on the approximations involved in using the simplest kinetic equations for the aggregation process. These include, for instance, ignoring multiplet mobility and multiplet-miltiplet collisions,
as well as the effect of mobility of aggregates (multiplets) on the diffusion constants used in the rate expressions, etc., especially in 
the beginning of the process when most aggregates are small. There are established methods in the literature that avoid some of such difficulties [21-34] and, in fact, several of the models for various reaction rates lead to particle-size distributions peaked and even sharpening with time [22,34,42,43]. 
However, the main point of our 
work, which we believe is new and crucial to obtaining narrow size
distributions, has been that the growth of the
secondary particles must be coupled, via the rate of generation of
singlets (primary particles), to the time-dependence of the process of 
formation of the latter; see the following sections.

Thus, we intentionally took the {\it simplest possible\/} models of both processes, the primary and secondary particle formation. This choice has the following advantages: it simplifies numerical simulations and thus 
allows to scan a wider range of model parameters; it avoids introduction
of unknown microscopic parameters. As a result, for instance, our description
of the aggregation process in this section has {\it no adjustable parameters};
they are all experimentally available. 
Similarly in the following sections, for primary particle formation, we only utilize {\it one adjustable parameter},
the effective surface tension, and even that parameter turns out to be
close to the experimental bulk value.

However, we recognize that more sophisticated modeling can
improve consistency with experiment, 
perhaps at the expense of additional assumptions and parameters, and we 
intend to explore this avenue
of investigation in future studies. In fact, we carried out preliminary numerical simulations allowing for the dimer diffusion and attachment to larger
aggregates and showing trend of improved consistency with the experiment.
We furthermore note that the approximation of restricting the aggregation
process to only the smallest particles sticking to the larger particles 
has been already used in the literature, e.g., [22,27]. 
Considerations of colloid stability have been typically utilized to justify
such approximations, and we note that detailed arguments of this sort would require additional microscopic parameters in the model.

We also comment that generally in dif\-fusion limited growth the aggregates
are expected to be fractal [38-41]. In this work we avoid
the issue of shape selection and internal structure of the secondary
particles; see a review [21]. There are likely some internal rearrangements and pushing out of the solvent going
on during the secondary particle growth for typical experimental conditions, because the particles are clearly spherical throughout the growth process. In some experiments [14] the surface of the secondary particles was initially ``hairy'' and
it got smoother at later times. We have assumed that the internal rearrangement processes are fast enough so that the shape and morphology of the aggregates are, respectively, spherical and compact. One could propose that  
the effects of internal rearrangements in our model make the 
coefficient in equation (3) another fit parameter rather than a fixed number. Distribution of the primary particle radii could also affect the porosity properties and thus modify this coefficient; we have not explored this matter.

\vskip 0.800 true cm

\noindent{\bf{}3.\ NUCLEATION AND GROWTH OF THE PRIMARY}\hfill\break
\noindent{\bf{}\hphantom{\bf 3.\ }PARTICLES}       
 
In this section we evaluate the primary-particle production rate, $\rho (t)$, assuming a fast nucleation process [8,9,34,44]. Thermodynamic models of ``burst'' nucleation of uniform dispersions have been originally formulated in [8,9]. Modeling the rate of formation of primary particles (singlets) requires setting up a master equation, where the rate of growth is determined by the Boltzmann factor with the thermodynamic free energy difference $\Delta G$, multiplied by $-1/(kT)$, in the exponent. This approach in turn requires modeling of the free energy of the growing embryos (sub-critical nuclei); in the simplest approach one can use the generic volume-plus-surface energy expressions.

Let us refer to the species (atoms, ions, molecules) which serve as monomers for the primary-particle nucleation as {\it solutes}. For a given  concentration $c(t)$ of solutes, larger than their equilibrium saturation concentration $c_0$ and approaching $c_0$ for large times $t$, the rate of formation
of critical nuclei can be written as [34,44]
$$ \rho(t)=4\pi a n_{cns}^{1/3} {\cal D} c^2 e^{-\Delta G_{cns} /kT} \, , \eqno(7) $$
\noindent which is based on the diffusional capture of solutes, whose effective radius is denoted by $a$, diffusion constant by ${\cal D}$, and $n$ is the number of solutes in an embryo. The subscript $cns$ refers to values calculated at the critical nucleus size. Note that $c(t)=n_1(t)$. 

The expression~(7) involves the following assumptions. For embryos of size $n < n_{cns}$, the solutes can be captured and emitted fast enough so that the
size distribution is given by the equilibrium form. Thus,
the factor $ce^{-\Delta G_{cns} /kT}$ in equation (7) follows
from the expectation that embryo sizes up to $n_{cns}$ are thermodynamically distributed, according the Boltzmann form. For sizes larger than $n_{cns}$, the dynamics is assumed to be fully irreversible, corresponding to an unbound growth by the capture of solutes.
The factor $4\pi a n_{cns}^{1/3} {\cal D} c$ in equation (7) is thus the appropriate version of the Smoluchowski growth rate similar to that in equations (2) and (3). The filling-fraction correction factor was absorbed in the definition of the effective solute radius $a$ to simplify the notation; it will be specified later.

For the free energy of the $n$-solute embryo, the following expression
will be used:
$$ G = - n kT \ln \left( c / c_0 \right) + 4 \pi a^2 n^{2/3} \sigma \, , \eqno(8) $$
\noindent which involves the bulk term, proportional to $n$, and the surface term.
The standard form of the bulk term was derived as follows. It is assumed that the entropic part of the free-energy change between the solid and solution phases can be calculated as the entropy in the supersaturated liquid suspension of solutes of concentration $c$, as given by the dilute (noninteracting) expression of the ``entropy of mixing,'' defined, e.g., in [45]. The surface term in equation (8) corresponds to the assumption that the growing embryos are spherical, of radius $a n^{1/3}$, and introduces their effective surface
tension $\sigma$, which is usually assumed to be comparable to the bulk surface tension.
It will be shown later that in the present case the results are very sensitive to the value of $\sigma$.
 
Obviously, all the above expressions are only meaningful for large $n$. It has been a common practice in the literature to use them for all $n$, as one of the approximations involved in a model. In what follows, we will in fact ignore
the difference between $G$ and $\Delta G = G(n)-G(1)$, and occasionally treat $n$ as a continuous nonnegative variable. 
Both  $n_{cns}$ and $\Delta G_{cns}$
are {\it explicit functions\/} of $c(t)$, 
$$ n_{cns}=\left[ 8 \pi a^2 \sigma \over 3 kT \ln \left(c/c_0\right) \right]^3 \, , \eqno(9) $$
$$ \Delta G_{cns} = {256 \pi^3 a^6 \sigma^3 \over 27 (kT)^2 \left[\ln \left(c/c_0\right) \right]^2}
\, , \eqno(10) $$
\noindent where the critical value $n_{cns}$ was calculated from $ \partial G/\partial n=0$.

Next, we account for the decrease in the concentration of solutes owing to the formation
of critical nuclei. Ordinarily, for $n > n_{cns}$ the primary particles grow (age) largely by absorbing diffusing solutes,  
and as in the preceding section we ignore here more complicated processes such as capture of small embryos, dissolution, etc. Simultaneously, the primary particles are also captured by the
secondary particles. In the present model, it is assumed for simplicity that the primary particles are captured fast enough by the growing secondary particles so that the effect of their aging on the concentration of solutes can be ignored. Furthermore, it has been generally recognized that aging, when significant, tends to sharpen the size distribution [21,34]. Thus, the primary particle radius $r$, introduced in the preceding section, will be assumed to have a single, experimentally determined value, although in reality [1] they have a finite-width, albeit not very wide, size distribution.
This approximation was already commented on earlier; it works largely because
only the product $rD$ matters in the rates in equations (1) through (3). 

Thus, we write
$$ {dc \over dt}=-n_{cns} \rho(t) \, , \eqno(11) $$
\noindent which means that the concentration of solutes is ``lost'' solely due to the irreversible formation of the critical-size nuclei. Collecting all the above expressions, one gets the following equations for $c(t)$ and $\rho(t)$:
$$ {dc \over dt}= - {16384 \pi^5 a^9 \sigma^4 {\cal D} c^2 \over 81 (kT)^4 \left[\ln \left(c/c_0\right) \right]^4}\,
\exp \left\{ -{256 \pi^3 a^6 \sigma^3 \over 27 (kT)^3 \left[\ln \left(c/c_0\right) \right]^2}
\right\} \, , \eqno(12) $$

$$ \rho (t) = {32 \pi^2 a^3 \sigma {\cal D} c^2 \over 3 kT \ln \left(c/c_0\right) }
\,
\exp \left\{ -{256 \pi^3 a^6 \sigma^3 \over 27 (kT)^3 \left[\ln \left(c/c_0\right) \right]^2}
\right\} \, .  \eqno(13)$$
It should be noted that replacing the distribution of the primary particle sizes by a single, experimentally measured average value of $r$, and ignoring their growth (ageing) after they achieve the critical size but before they are captured, violates the conservation of matter. Thus, {\it in the present model only the shape of the secondary particle size distribution is relevant}. The absolute number densities $N_s$ must be rescaled to correspond to the actual amount of the solid matter per unit volume. The latter data are usually available experimentally.
 
\vskip 0.800 true cm

\noindent {\bf 4.\ RESULTS AND COMPARISON WITH AN EXPERIMENT}

In order to test the model developed in Sections 2 and 3, we used experimental data
on the growth of dispersions of submicron spherical 
gold particles [1,7], which were produced by the reduction of chloroauric acid 
(HAuCl$_4$) with ascorbic acid.  The simplicity of this system and the possibility to either measure or estimate all the 
necessary parameters make it ideally suited for testing the theoretical model.
In synthesis in concentrated dispersions, the aggregation process is assured, 
resulting in the formation of particles consisting of a large number of subunits. The 
precipitation procedure used in [1,7] has resulted in spherical 
gold particles of a narrow size distribution. After a short induction period
(up to 6-8 sec) nucleation occurs, followed by immediate aggregation. The total process time varies from 3 to 20 sec, depending on the experimental conditions selected.
Scanning electron micrographs of the resulting particles can be seen in [1,7].

The field emission microscopy [1,7] revealed the presence of the 
subunits, having an approximate size of 30 to 40$\,$nm. The 
calculated value for the packing fraction of the subunits in the 
aggregated secondary particles was characteristic of a random loose packing, 
usually expected in the formation of rapidly assembled systems [36].  The size of the 
primary particles was estimated from the X-ray diffraction measurements.  Calculations have generated values between 30 and 42 nm, which were in excellent 
agreement with the subunit size data from electron microscopy. Information on the size
distribution of the primary particles is also available [1,7].

Numerical simulations
required to follow the time evolution of the kinetic equations turned out to be large-scale. Therefore, the testing of the model was restricted to one randomly selected set of experimental parameters. 
However, we actually varied numerically all the
parameter values and found that the calculation results are affected to various degree by them. The parameter values will be discussed roughly in the order of increasing sensitivity of the numerical results.
For consistency all input data are in the MKS system of units.

The radius of the primary particles,
$$ r=2.10 \cdot 10^{-8} \, {\rm m} \, , \eqno(14) $$
\noindent was obtained experimentally, as described in the preceding section, which is
within the range of $0.5\cdot 10^{-8} \, {\rm m}$ to
$5\cdot 10^{-8} \, {\rm m}$, typical for the system under consideration. 
The value in equation (14) was for the experiment [1] for which the initial concentration $c(0)$ was
$$ c(0)=6.0 \cdot 10^{25} \, {\rm m}^{-3} \, ; \eqno(15) $$
\noindent $ c(0)$
was calculated from the concentration of the gold solution used in the preparation of the dispersion, and it yields the initial condition in equations (12)-(13). The diffusion constant $D$ of the primary particles was obtained from the Einstein formula,
$$ D=1.03 \cdot 10^{-11} \, {\rm m}^2\, {\rm sec}^{-1} \, , \eqno(16) $$
\noindent with
$$ kT=4.04 \cdot 10^{-21} \, {\rm J} \, . \eqno(17) $$

The saturation concentration of gold in solution, $c_0$, is not well known and is expected to depend somewhat on the experimental conditions.
Using $2\cdot 10^{-12}\,$mol$\,$dm$^{-3}$ \ [46] yields 
$$c_0 \simeq 1\cdot 10^{15} \, {\rm m}^{-3} \, . \eqno(18) $$
\noindent The results for the particle size distribution are not particularly sensitive to this parameter, because it enters under logarithm in equations (12)-(13). The solute diffusion constant, 
$$ {\cal D}=1.5 \cdot 10^{-9} \, {\rm m}^2 \, {\rm sec}^{-1} \, , \eqno(19) $$
\noindent was estimated similarly to $D$ in equation (16), using the
Einstein formula, with the radius
of the gold atom of $1.44 \cdot 10^{-10} \,$m [47,48].
The applicability of this formula to single atoms may not be exact, but the particle size distribution is not too sensitive
to this parameter value: a decrease in ${\cal D}$ shifts the calculated distribution to somewhat smaller aggregate sizes.

It was established numerically that the size distribution of the secondary particles was sensitive to the values of $a$, the effective atomic (solute) radius,
and to the surface tension $\sigma$.
Note that $a$ was defined to relate
the number of solutes $n$ in a growing primary particle to its radius, given by $an^{1/3}$. It is assumed that the primary particles are largely crystalline; thus, the best choice of $a$ is such that $4\pi a^3
/3$ is the volume per atom, including the attributable part of the surrounding void volume, in bulk gold. Consequently,
$$ a=1.59 \cdot 10^{-10} \, {\rm m} \,  \eqno(20) $$
\noindent was obtained by dividing the radius of the gold atom
($1.44 \cdot 10^{-10} \, {\rm m}$) by the cubic root of the volume filling fraction, 0.74, of the crystalline structure of gold [36]. 

The effective surface tension of nanosize gold embryos in solution, $\sigma$, profoundly affects the numerical results. Unfortunately, even the bulk-gold value, which is of order
$$ \sigma \simeq 0.58 \; {\rm to}\; 1.02 \; {\rm N}/{\rm m} \, , \eqno(21) $$
\noindent is not well known [49], and it may differ from that of the nanosize solids. Given
this fact, $\sigma$ was chosen as the only adjustable
parameter in the model.

Experimentally, the time scale on which the secondary particle growth effectively terminated was about 8 to $10 \, {\rm sec}$, which does not include the ``induction'' stage. In Figures 1 though 3, the results of the numerical simulations of the kinetic equations are presented with parameters as specified above, for three different values
of $\sigma$, which clearly demonstrate the sensitivity to the choice of this parameter. 
In Figure 1, the case $\sigma=0.51\,$N$/$m illustrates growth that already reached saturation
for times up to $10 \, {\rm sec}$. It should be noted that the distribution evolves quite slowly with time. Initially, it is heavily weighed in the small-aggregate regime. Later on, the large-size peak develops and eventually dominates the distribution.
By varying $\sigma$ near the expected range, given in equation (21), it was found that, for times up to $10 \, {\rm sec}$, {\it all\/} $\sigma$ values yielded smaller average sizes than the experimentally measured one,
$$ \left( R_s \right)_{\rm average\ (experimental)} = 1.0 \pm 0.1 \, \mu{\rm m} \, . \eqno(22) $$
\noindent Seeking $\sigma$ that would give the largest sec\-ondary particle size resulted in 
$$ \left( \sigma \right)_{\rm fitted} =  0.57 \pm 0.04 \, {\rm N}/{\rm m}
 \, , \eqno(23) $$
\noindent which agrees well with the bulk value in equation (21).

Figure 2 shows the size distribution for $\sigma=0.57\,$N$/$m. The growth did not reach the full saturation at the relevant times, and
the peak particle radius at $t=10 \, {\rm sec}$, of $R_s \simeq 0.32 \, \mu{\rm m}$, is smaller than the experimental value in equation (22). The width of the distribution, of $\sim$10$\,$\%{}, is close to that established experimentally. Given the approximations involved in the model, only semiquantitative agreement with the experimental data should be expected. Since the key feature of the model is the prediction of the narrow-width distribution of secondary particle sizes, the overall consistency with the experimental results is gratifying.

As the value of $\sigma$ is increased, the large-size peak does not fully develop on the relevant time scales, as exemplified by the case $\sigma=0.63\,$N$/$m shown in Figure 3. The reader should be reminded that, owing to the absence of the conservation of matter in this model, the number densities $N_s$ must be rescaled according to the actual amount of the solid matter per unit volume, if the comparison of the calculated and experimental distribution is attempted.

Asymptotically, the particle-size distribution
``freezes'' for large times, i.e., the particle growth actually stops in this model as opposed to the self-similar growth studied, for instance, in [37]. Equation (11), with (9), can be integrated in closed form to yield
$$ \int\limits_0^t \rho (t^\prime) \, d t^\prime = \left( { 3kT
\over 8 \pi a^2 \sigma } \right)^3 c_0 \, \left[ F\left( {c(0)
\over c_0 } \right) - F\left( {c(t)
\over c_0 } \right) \right] \, , \eqno(24) $$
\noindent where 
$$ {F(x)\over x}=( \ln x )^3 - 3 ( \ln x )^2
+6 ( \ln x ) -6  \, . \eqno(25) $$
\noindent The left-hand side of equation (24) is the total number of primary particles produced by the time $t$. From equations (24)-(25), this number is finite as $t \to \infty$, when $c(t) \to c_0$. The supply of the primary particles, manifested by the peak at small sizes for short times,
in Figures 1-3, is essential for the large-size peak in the distribution to develop and grow, because the present model assumes the growth of the secondary particles to be solely by singlet capture.

Let us consider the quantity $\tau$ defined [1] by
$$ \tau = \int\limits_0^t 
{N}_1(t^\prime)\, dt^\prime \, . \eqno(26) $$
\noindent If the independent variable is changed from $t$ to $\tau$
in equations (1)-(4), the resulting relations are linear in $N_{s>1}$. One can establish that the only way to have a normalizable stationary size-distribution as $t \to \infty$
is to have $\tau (t)$ approach a {\it finite\/} value for large $t$. This quantity was calculated numerically for the $\sigma$ values used in Figures
1-3. The results, shown in Figure~4, confirm the earlier observation that the growth process saturates fast for $\sigma=0.51\,$N$/$m. For the two larger $\sigma$ values there is still some variation for the time scales of order 1 to 10$\,$sec; see Figure 4. The function $\tau(t)$ is useful in identifying the time scales of the growth process. 

In summary, we reviewed a new model explaining synthesis of submicron size polycristalline colloid particles with the size distribution that is peaked at an average value corresponding to a large number of primary particles in a final secondary particle. For the experimental gold-sol system the model has worked reasonably well: the average size, the width of the distribution, the time scale of the process, and even the fitted effective surface tension were all semiquantitatively consistent with the measured or expected values. The present model has involved several simplifying assumptions. Future studies will incorporate additional effects in the model and test it for a wider range of experimental systems. The main conclusion has been that multistage growth models can yield size-selection as a kinetic phenomenon, which has been observed in a large number of experimental systems.

\vfill\eject

\centerline{\bf REFERENCES}{\frenchspacing{}\raggedright{}

\item {1.} \parskip 2 pt{}Privman, V., Goia, D. V., Park, J., and Matijevi{\' c}, E. 
1999, J. Colloid Interf. Sci. {\bf 213}, 36.

\item {2.} Matijevi{\' c}, E. 1985, 
Ann. Rev. Mater. Sci. {\bf 15}, 483.

\item {3.} Haruta, M., and Delmon, B. 1986, 
J. Chim. Phys. {\bf 83}, 859.

\item {4.} Sugimoto, T. 1987, 
Adv. Colloid Interf. Sci. {\bf 28}, 65.

\item {5.} Sugimoto, T. 1982, 
J. Colloid Interf. Sci. {\bf 150}, 208.

\item {6.} Matijevi{\' c}, E. 1994, 
Langmuir {\bf 10}, 8.

\item{7.} Goia, D. V., and Matijevi{\' c}, E. 1999,
Colloids Surf. {\bf 146}, 139.
 
\item{8.} LaMer, V. K. 1952, 
Ind. Eng. Chem. {\bf 44}, 1270.

\item{9.} LaMer, V. K., and Dinegar, R. H. 1950, 
J. Amer. Chem. Soc. {\bf 72}, 4847.

\item{10.} Murphy-Wilhelmy, D., and Matijevi{\' c}, E. 1984, 
J. Chem. Soc., Faraday Trans. I {\bf 80}, 563.

\item{11.} Matijevi{\' c}, E., and Murphy-Wilhelmy, D. 1982, 
J. Colloid Interf. Sci. {\bf 86}, 476.

\item{12.} Matijevi{\' c}, E., and Scheiner, P. 1978, 
J. Colloid Interf. Sci. {\bf 63}, 509.

\item{13.} Hsu, U. P., R\"onnquist, L., and Matijevi{\' c}, E. 1988, 
Langmuir {\bf 4}, 31.
 
\item{14.} Oca\~na, M., and Matijevi{\' c}, E. 1990, 
J. Mater. Res. {\bf 5}, 1083.

\item{15.} Oca\~na, M., Serna, C. J., and Matijevi{\' c}, E. 1995, 
Colloid Polymer Sci. {\bf 273}, 681.

\item{16.} Lee, S.-H., Her, Y.-S., and Matijevi{\' c}, E. 1997,
J. Colloid Interf. Sci. {\bf 186}, 193.

\item{17.} Edelson, L. H., and Glaeser, A. M. 1988,
J. Am. Chem. Soc. {\bf 71}, 225.

\item{18.} Bailey, J. K., Brinker, C. J., and Mecartney, M. L. 1993,
J. Colloid Interf. Sci. {\bf 157}, 1.

\item{19.} Morales, M. P., Gonz\'ales-Carre\~no, T., and Serna, C. J. 1992, 
J. Mater. Res. {\bf 7}, 2538.

\item{20.} Oca\~na, M., Morales, M. P., and Serna, C. J. 1995,
J. Colloid Interf. Sci. {\bf 171}, 85.

\item{21.} Dirksen, J. A., and Ring,  T. A. 1991,
Chem. Eng. Sci. {\bf 46}, 2389.

\item{22.} Dirksen, J. A., Benjelloun, S., and Ring, T. A. 1990,
Colloid Polym. Sci. {\bf 268}, 864.

\item{23.} Ring, T. A. 1991, Powder Technol. {\bf 65}, 195.

\item{24.} van Blaaderen, A., van Geest, J., and Vrij, A. 1992,
J. Colloid Interf. Sci. {\bf 154}, 481.

\item{25.} Bogush, G. H., and Zukoski, C. F. 1991, 
J. Colloid Interf. Sci. {\bf 142}, 1 and 19.

\item{26.} Lee, K., Look, J.-L., Harris, M. T., and McCormick, A. V. 1997,
J. Colloid Interf. Sci. {\bf 194}, 78.

\item{27.} Look, J.-L., Bogush, G. H., and Zukoski, C. F. 1990,
Faraday \hfill\break{}Discuss. Chem. Soc. {\bf 90}, 345 and 377.

\item{28.} Randolph, A. D., and Larson, M. A. 1988, ``Theory of 
Particulate Processes'' (Academic Press, San Diego).

\item{29.} Brinker, C. J., and Scherer, G. W. 1990, ''Sol-Gel Science''
(Academic Press, Bos\-ton).

\item{30.} Flagan, R. C. 1988, Ceramic Trans. {\bf 1} (A), 229.

\item{31.} Scott, W. T. 1968, J. Atmospheric Sci. {\bf 25},
54.

\item{32.} Higashitani, K. 1979, J. Chem. Eng. Japan
{\bf 12}, 460.

\item{33.} Ludwig, F.-P., and Schmelzer,  J. 1996,
J. Colloid Interf. Sci. {\bf 181}, 503.

\item{34.} Overbeek, J. Th. G. 1982,
Adv. Colloid Interf. Sci. {\bf 15}, 251.

\item{35.} Weiss,  G. H. 1986,
J. Statist. Phys. {\bf 42}, 3.

\item{36.} German, R. M. 1989, ``Particle Packing Characteristics'' 
(Metal Powder Industries Federation, Princeton).

\item{37.} Brilliantov, N. V., and Krapivsky, P. L. 1991,
J. Phys. A {\bf 24}, 4787.

\item{38.} Krug, J., and Spohn, H. 1991, in ``Solids Far from 
Equilibrium,'' edited by Godrech{\` e}, C. (Cambridge 
University Press).

\item{39.} Family, F., and 
Vicsek, T. 1991, ``Dynamics of Fractal 
Surfaces'' (World Scientific, Singapore).

\item{40.} Medina, E., Hwa, T., Kardar, M., and Zhang, Y.-C. 1989,
Phys. Rev. A {\bf 39}, 3053.

\item{41.} Schaefer, D. W., Martin, J. E., Wiltzius, P., and
Cannell, D. S. 1984, Phys. Rev. Lett. {\bf 52}, 2371.

\item{42.} Reiss, H. 1951, J. Chem. Phys. {\bf 19}, 482.

\item{43.} Mozyrsky, D., and Privman, V. 1999, J. Chem. Phys.
{\bf 110}, 9254.

\item{44.} Kelton, K. F., Greer, A. L., and Thompson, C. V. 1983,
J. Chem. Phys. {\bf 79}, 6261.

\item{45.} Reif, F. 1965, ``Fundamentals of Statistical and Thermal 
Physics'' (McGraw-Hill, New York).

\item{46.} Linke, W. F. 1958, ``Solubilities of Inorganic and Metal-Organic
Compounds,'' 4$^{\rm th}$ ed., Vol. 1, p. 243 (Van Nostrand, 
Princeton).

\item{47.} Cotton, F. A., Wilkinson, G. and Gauss, P. L. 1995,
``Basic Inorganic Chemistry,'' 3$^{\rm rd}$ ed., p. 61 (Wiley, 
New York).

\item{48.} ``Gmelins Handbuch der Anorganischen 
Chemie,'' 1954, 8$^{\rm th}$ ed., p. 429 (Verlag Chemie, Weinheim).

\item{49.} ``American Institute of Physics 
Handbook'' 1957, edited by Gray, D. E., 3$^{\rm rd}$ ed., p. 2-208.

}

\pageinsert

%\hphantom{A}\vskip 11.8 true cm

\noindent\epsfig{file=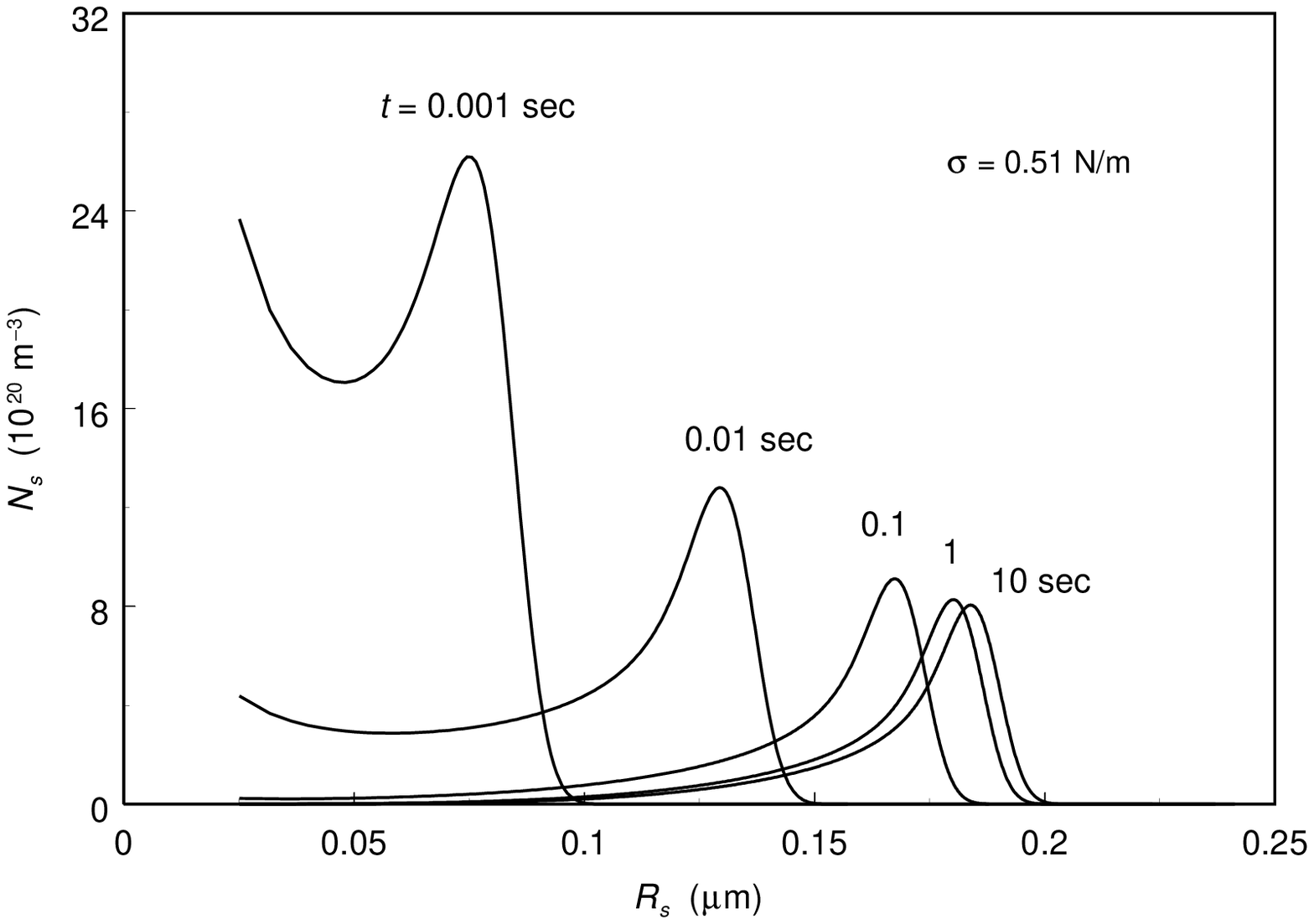,width=11cm}

\noindent Figure~1:\ \ Distribution of the secondary particles by their sizes, calculated\hfill\break 
\hphantom{Figure~1:\ \ }for times $t=0.001$, 0.01, 0.1, 1, 10 sec, using $\sigma=0.51\,$N$/$m.

\endinsert

\pageinsert

%\hphantom{A}\vskip 11.8 true cm

\noindent\epsfig{file=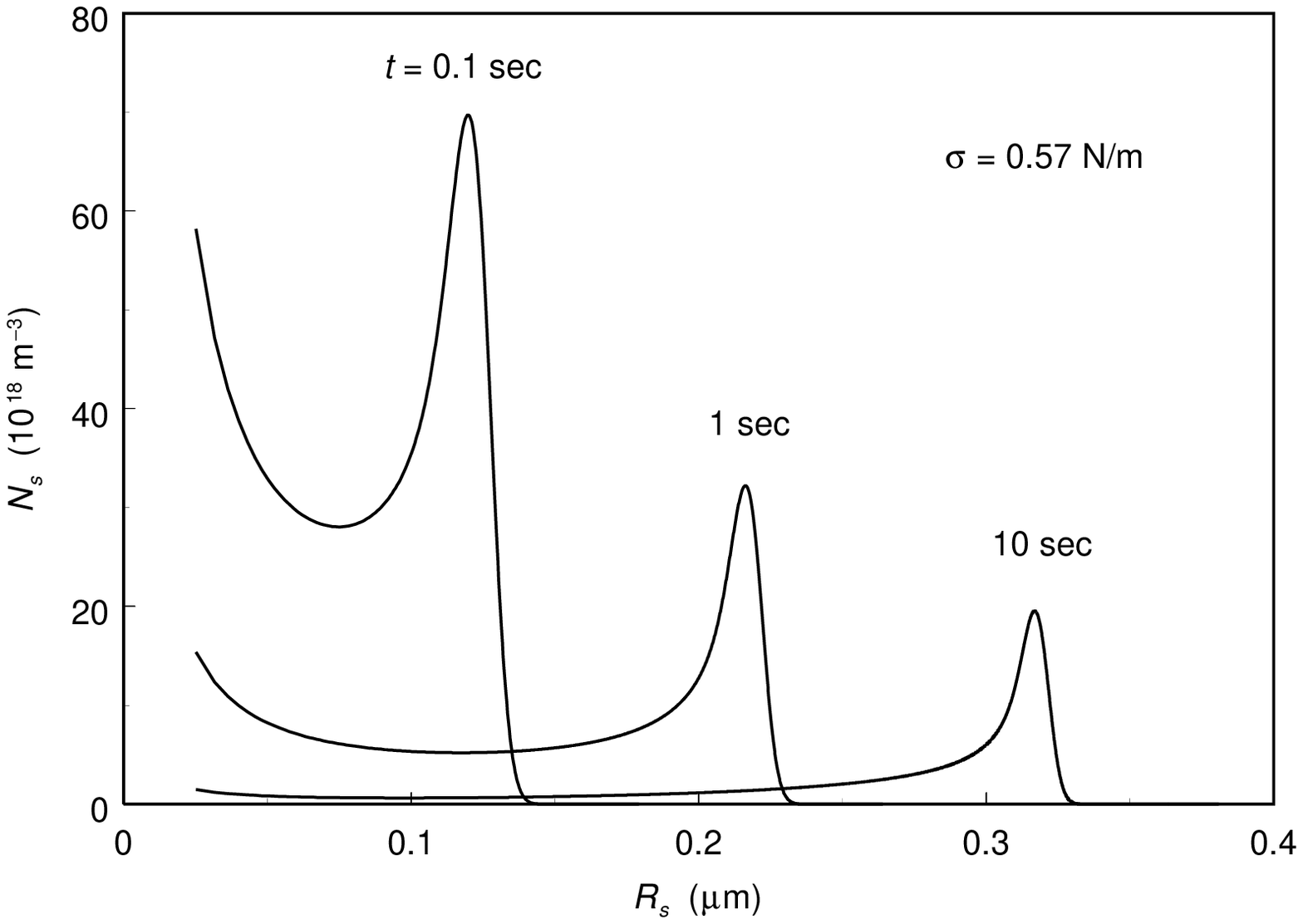,width=11cm}

\noindent Figure~2:\ \ The same plot as in Figure~1, but using $\sigma=0.57\,$N$/$m,
for times\hfill\break 
\hphantom{Figure~2:\ \ }$t=0.1$, 1, 10 sec.
\endinsert

\pageinsert

%\hphantom{A}\vskip 11.8 true cm

\noindent\epsfig{file=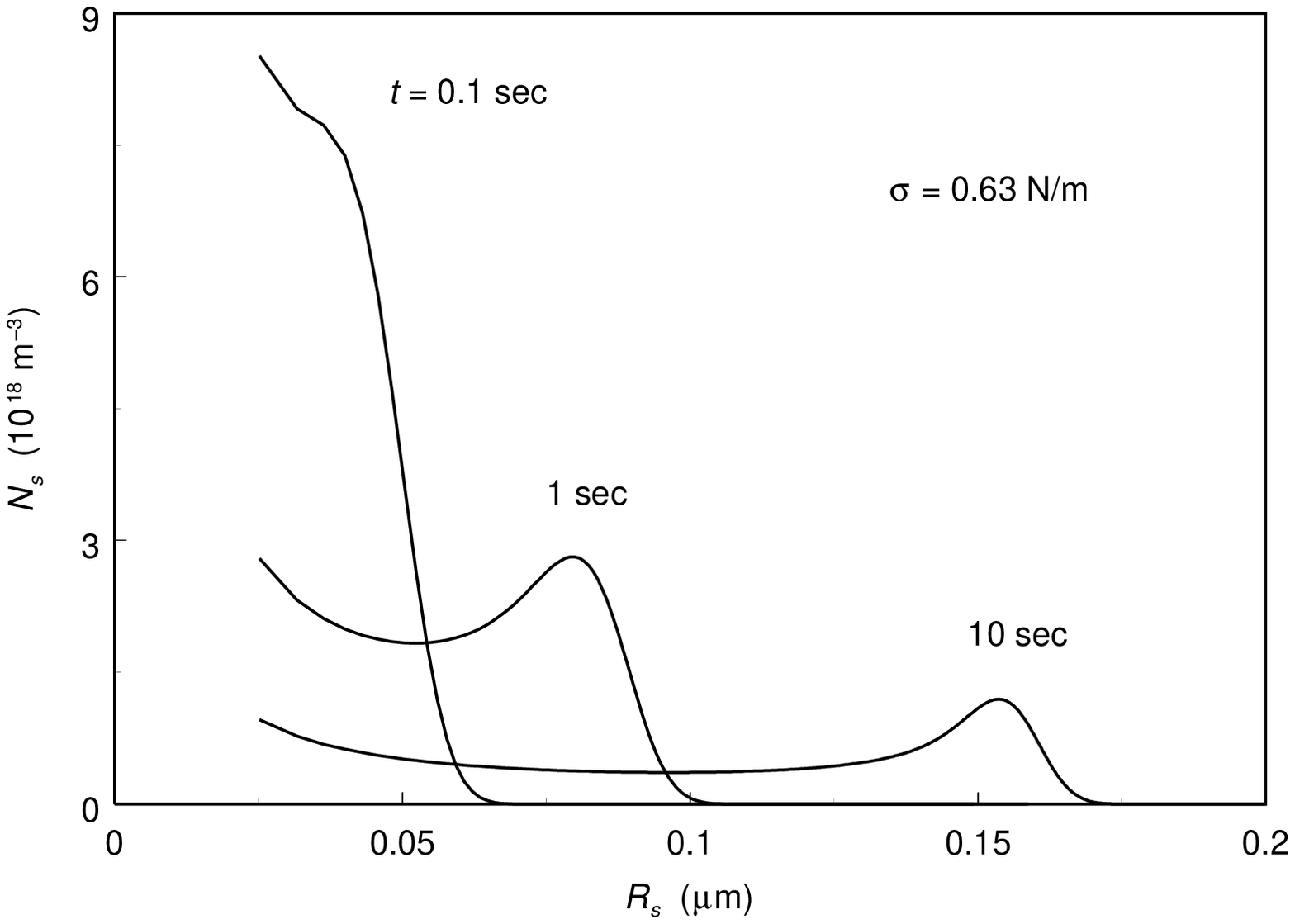,width=11cm}

\noindent Figure~3:\ \ The same plot as in Figure~1, but using $\sigma=0.63\,$N$/$m,
for times\hfill\break 
\hphantom{Figure~3:\ \ }$t=0.1$, 1, 10 sec.

\endinsert{}

\pageinsert

%\hphantom{A}\vskip 11.8 true cm

\noindent\epsfig{file=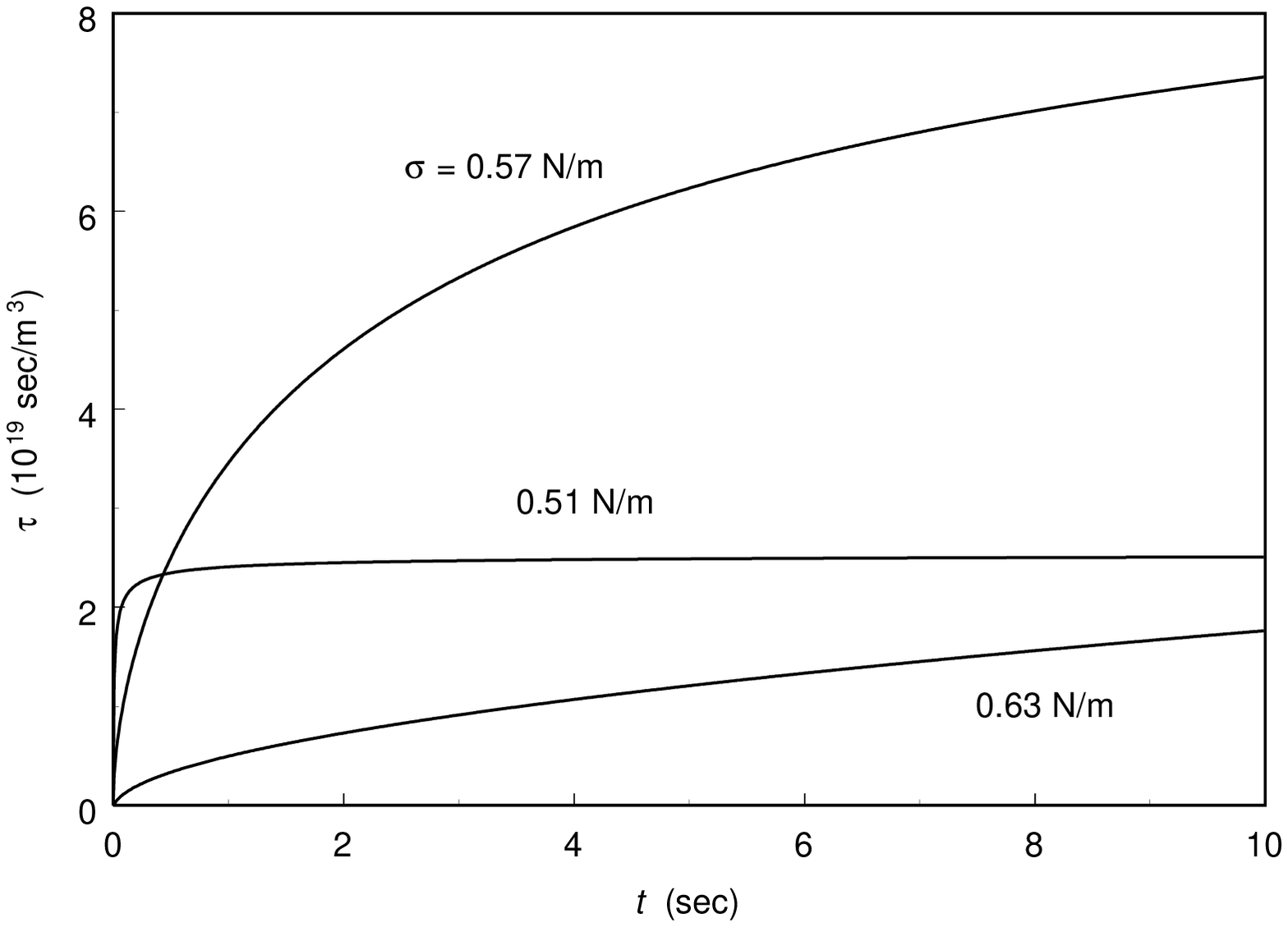,width=11cm}

\noindent Figure~4:\ \ The function $\tau (t)$ calculated for the values of $\sigma$ corresponding\hfill\break 
\hphantom{Figure~4:\ \ }to Figures 1-3.
\endinsert{}

\bye